# UV production of methane

# from surface and sedimenting IDPs on Mars

# in light of REMS data and with insights for TGO

Corresponding Author:

John E. Moores

Center for Research in Earth and Space Science (CRESS),

York University, 4700 Keele Street, Toronto, ON M3J 1P3, Canada  jmoores@yorku.ca

Contributing Authors:

Christina L. Smith: CRESS, York University, chrsmith@yorku.ca

Andrew C. Schuerger: Dept. of Plant Pathology, University of Florida, USA,

schuerg@ufl.edu

## Abstract

This paper refines model predictions for the production of methane from UV-irradiated interplanetary dust particles (IDPs) now that the Rover Environmental Monitoring Station (REMS) instrument onboard the Mars Science Laboratory (MSL) Rover has made the first measurements of the UV environment on the surface of Mars, at Gale Crater. Once these measurements are included in a UV radiative transfer model, we find that modelled UV sol-integrated energies across the planet are lower than pre-measurement estimates by 35% on average, considering all latitudes and seasons. This reduction, in turn, reduces the predicted production of methane from individual accreting IDPs, extending their lifetimes and increasing the surface concentration of organics that must accumulate in order to emit sufficient methane to balance the accretion of organic compounds to Mars. Emission from reasonable accumulations of IDPs could range up to ~$7.9 \times 10^{-4}$ ppbv sol$^{-1}$. Richer deposits of organic carbon at the surface may emit methane at no more than 3.9 ppbv sol$^{-1}$. An examination of IDP-derived methane production during atmospheric settling indicates that no more than 0.32% of organic carbon from meteor streams may be deposited in the atmosphere. Thus, such a process cannot explain either the spikes observed in methane nor the low equilibrium values observed by MSL. Instead, this discrepancy may be explained if < 80 tons per year of organic carbon survives to the surface, the atmospheric lifetime of methane is < 110 years or the efficiency of the UV-CH$_4$ process is < 7%. Under the assumption of reduced carbon input cycling in the Martian system from these processes, both soil concentrations of organic carbon and atmospheric measurements of methane observed by MSL are consistent with the UV-CH$_4$ process. This refinement of methane production from IDPs and its geographical and vertical distribution will be an important input for models attempting to understand the results to be derived from the Trace Gas Orbiter (TGO) mission that will map methane concentrations in the martian atmosphere in 2018 at 0.01 ppbv.

***Index Terms: Mars, methane, atmospheric chemistry, UV***

# 1. Introduction

Laboratory studies (Stoker and Bullock, 1998; Schuerger et al., 2012; Keppler et al., 2012) demonstrate that methane is readily evolved from interplanetary sources of organic carbon when irradiated by ultraviolet (UV) photons between 200 and 400 nm. Furthermore, Mars should continuously accrete such particles, as dynamical modelling of interplanetary dust particles (IDPs) in the plane of the solar system suggests that several hundred tons of this organic carbon should be collected by the upper atmosphere of Mars each year (Flynn, 1996). It was this combination which led to the development of a UV-$CH_4$ model that linked together modelled UV irradiation of Mars (e.g. Moores et al., 2007) with the input of organic carbon to calculate the surface loading of organics (Moores and Schuerger, 2012) and the equilibrium methane concentration in the atmosphere (Schuerger et al. 2012). This modelling found that if the entire input of organic carbon was converted to methane with an atmospheric lifetime of 329 years (Atreya et al., 2007), 11 ppbv would accumulate as a steady state concentration over geological time. However, a more reasonable upper bound of 2.2 ppbv based upon the IDP H/C ratio was considered more likely (Schuerger et al. 2012). These values were roughly consistent with many of the telescopic and orbital measurements that were made prior to the landing of the Mars Science Laboratory (MSL) rover in late 2012.

The MSL rover carried two instruments capable of providing information to test the UV-$CH_4$ model: the Sample Analysis at Mars Tunable Laser Spectrometer (SAM-TLS) which would directly measure the atmospheric concentration of methane to a precision of ±0.1 ppbv (Webster and Mahaffy, 2011; Mahaffy et al., 2012) and the Rover Environmental Monitoring Station (REMS) which would quantify, for the first time, the amount of UV radiation received at the surface of Mars (Gómez-Elvira et al., 2012). Following initial null results (Webster et al., 2013) that limited methane values below 1 ppbv, SAM-TLS found evidence for large spikes of methane up to ~7 ppbv which persisted over relatively short timescales of tens of sols and a background concentration of methane of < 0.7 ppbv (Webster et al., 2015). This value was more than 3 times lower than predicted by the UV-$CH_4$ model and is less than can be explained by the discrepancy between pre-flight modelling of UV flux (Moores et al., 2007) and the observed UV flux (Smith et al., 2016). This disagreement is particularly interesting in light of the organic content of soils at Gale crater (Freissinet et al., 2015) which are lower than anticipated, but lie within pre-flight predictions made using the UV-$CH_4$ model (Moores and Schuerger, 2012), consistent with all other landing sites yet visited on Mars.

What could be causing the disagreement for atmospheric methane between the UV-$CH_4$ model and the SAM-TLS results? One possibility is that the effective lifetime of methane in the Martian atmosphere is significantly less than 329 years. A low atmospheric lifetime for methane, perhaps due to an unknown destruction mechanism, was cited by Lefèvre and Forget (2009) as a necessary condition to explain the rapid disappearance of the methane plumes reported by Mumma et al. (2009). Another possibility is that relatively little organic carbon reaches the surface in the first place (Fries et al., 2016), due to photolysis of falling IDPs high in the atmosphere yielding methane that is immediately destroyed by the high fluxes of lyman-$\alpha$ radiation above 70 km in altitude (Wong et al., 2003). Thirdly, the efficacy of the UV-CH$_4$ process could be less than 20% (Schuerger et al., 2012). Finally, there are mechanisms that cause methane-cycle compounds to be redistributed across the surface, including the aeolian transport of small particles that would include IDPs (Moores and Schuerger, 2012), atmospheric circulation (Mischna et al., 2011), and adsorption and desorption of methane onto regolith (Gough et al., 2010; Meslin et al., 2011). Each of these processes may affect the observed local concentration of methane.

However, neither of these possibilities fundamentally challenges the UV-CH$_4$ production process. Given the imminent observation of methane with a precision of $\pm$ 10 pptv (0.01 ppbv) by the Mars Trace Gas Orbiter (TGO) in 2018 (Robert et al., 2016), it is necessary to refine the UV-CH$_4$ model using the REMS results to provide information on UV-mediated production of methane. However, note that this paper focuses solely on production and will not discuss the redistribution of methane in the atmosphere via circulation. The production of methane will be a critical input into models which contain destruction mechanisms and which can then inform the interpretation of the TGO methane results. Furthermore, as noted by Moores and Schuerger (2012), measurements of methane emissions from the surface can be used to constrain the quantity and distribution of organic carbon at the surface of Mars. As such, this paper provides the needed update to the model, the specifics of which will be discussed in section 2. Section 3 will then describe the refined UV-CH$_4$ production as applied to Mars. Finally, this refinement will be used as a framework to critically evaluate the organic carbon budget of Mars including the proposed carbon inputs, methane destruction mechanisms and their likely geographic and temporal variation (sections 4.1 and 4.2) before providing specific implications for TGO observations (4.4). The question of aerial deposition of methane from sedimenting IDPs will also be considered in section 4.3.

## 2. Materials and Methods

### 2.1 Updating the Tomasko Model using REMS

#### 2.1.1 Input Parameters

The UV simulations use the Doubling and Adding (D&A) radiative transfer code of Griffith et al. (2012), adapted for Martian conditions. The Griffith et al. (2012) model, like the Moores et al. (2007) model used in previous work (Moores and Schuerger, 2012) traces its heritage back to the Martian D&A model of Tomasko et al., (1999). As such, they are functionally identical. The D&A model used a two-layer/three-level configuration and included gaseous absorption, Rayleigh scattering and Mie scattering from a variety of species. The upper level was assumed to contain all the gaseous absorption and some Rayleigh scattering. The lower level was assumed to contain all the Mie scattering centers and some Rayleigh scattering.

The UV wavelength range considered in this model was 200-400 nm. Contributions to the flux received by the ground at wavelengths between 100-200 nm are negligible due to very high absorption by $CO_2$ (Kuhn and Atreya, 1979). As the optical depth contributions due to gaseous absorption and Rayleigh and Mie scattering are wavelength dependent, the wavelength range was split into six different regions, henceforth referred to as "bands": UVA (315-400 nm), UVB (280-315 nm), and four UVC bands (i: 200-220 nm, ii: 220-240 nm, iii: 240-260 nm, and iv: 260-280 nm).

The optical depths due to Rayleigh scattering, $\tau_R$, were computed following the relation of Hansen and Travis (1974):

$$\tau_R = \frac{P\sigma}{g\mu} = \frac{P(6+3\delta)}{g\mu(6-7\delta)}\frac{8\pi^3}{3N^2\lambda^4}\left(n_g^2-1\right)^2 \qquad [\,1\,]$$

where $P$ is the pressure, $g$ is the surface gravity of the planet (3.71 ms$^{-2}$ for Mars), $\mu$ is the mean molecular mass of the gas in question, $\sigma$ is the Rayleigh scattering coefficient as a function of wavelength, $n_g$ is the refractive index as a function of wavelength, $N$ is Lorschmidt's number when, as in this work, the refractive indices are computed at standard temperature and pressure, and $\delta$ is the depolarization parameter. The refractive indices were taken from Cox et al. (2001), and the depolarization factors from Hansen and Travis (1974) and Penndorf et al. (1957). The optical depths were computed over a fine wavelength grid across the 200-400 nm range. The values obtained within a particular band were averaged to produce an average optical depth due to scattering across the entire band. The species included were: $CO_2$ (95%), $N_2$ (2.7%), Ar (1.6%), and $O_2$ (0.13%) and the total Rayleigh scattering was given by:

$$\tau_R = \sum_i \tau_{Ri} f_i \qquad [\,2\,]$$

where $f_i$ is the mixing ratio of species $i$. Gaseous absorption was considered for $CO_2$ and $O_2$. The optical depth contributions from gaseous absorption for each species, $\tau_a$, were calculated as averages across each band according to:

$$\tau_a = \acute{\sigma}\,n \qquad [\,3\,]$$

where $\acute{\sigma}$ is the average absorption cross-section across the band being considered. The cross-sections as functions of wavelength were taken from the literature. For $CO_2$: Shemansky (1972), for $O_2$: Ackerman et al. (1971) and Bogumil et al. (2003) via Keller-Rudek et al. (2013) from experiments at approximately 298 K. The column abundances, $n$ were taken as $CO_2$: 74 x$10^6$ μm-atm, $O_2$: 104 μm-atm. The absorption cross-sections for each species vary significantly across the UVC band, hence this band was subdivided into four segments. In spectral regions in which no cross-sectional data were available, the cross-sections were assumed to be zero. This assumption was made as it was observed in each of these regions that the cross-sections tended towards zero.

Furthermore, it was assumed that the gaseous optical depths and Rayleigh scattering optical depths did not vary with position or solar longitude, and thus were kept constant in all models.

The aerosols were assumed to be comprised of dust particles that are well described by Mie scattering (Smith et al. 2016). The Mie scattering centers were assumed to be cylindrical particles with a one-to-one length-to-diameter relation, as per the published parameters of Smith et al. (2016) and Wolff et al. (2010). Non-spherical Mie scattering centres are included in the D&A code using the empirical method of Pollack and Cuzzi (1980), requiring the specification of four additional parameters: the surface area ratio (calculated from the cylindrical particle assumption), the scattering angle at which the log of the phase-function is at minimum (taken from the adopted Band 6 phase function of Wolff et al. 2010), a boundary parameter defining the particle size limit below which Mie scattering is a good representation of the scattering (as per previous works, e.g. Tomasko et al. 1999, this was assumed to be 5 μm) and an empirical constant related to the slope of the phase function- scattering angle curve at a scattering angle of zero degrees (taken directly from Tomasko et al. 1999 for 443nm - at small scattering angles, the phase functions of Tomasko et al. 1999 are in agreement with that of Wolff et al. 2010). The effective radius was 1.4μm and the effective variance was 0.3. The Mie particle parameters for the UVA and UVB bands were taken as those of band 7 parameters from Wolff et al. (2010) and UVC were taken as band 6 parameters, case 1 (no ozone), from Wolff et al. (2010). The Mie scattering center abundances varied with model.

The ground is assumed to be a Hapke surface with parameters taken from Smith et al. (2016). These were taken as constant across all bands: $b_0 = 0.8$, $h = 0.06$, $c = 0.45$, $b = 0.30$. However, $w$ varied with wavelength and was taken as 0.0095 (UVA, UVB) and 0.0070 (UVC) as in Wolff et al. (2010). Each of these parameters is defined as per the notation of Johnson et al. (2006).

*2.1.2 Model Validation*

The model was validated by comparing the output of this model with that of Figure 3 of Smith et al. (2016). The model uses the total incident UVA flux on a flat surface as a function of optical depth (inputted directly). The Solar zenith angle was taken as 15º, and the input Solar flux at the top of the atmosphere in the relevant wavelength band was taken as 21.0 $Wm^{-2}$, consistent with the input parameters used to produce this figure (Smith, 2017 - personal communication). The diffuse and direct downward UV fluxes and the net downward flux (direct+diffuse-upwards) were compared at the base of the second layer. The results are shown in Figure 1.

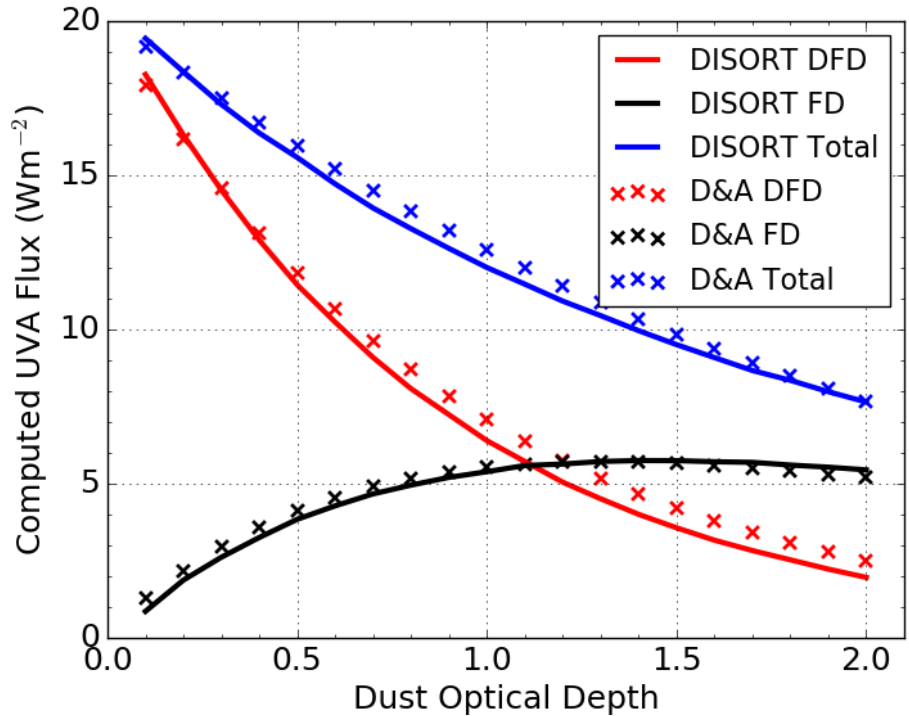

**Figure 1** - *Model results of this study (crosses) in comparison to those of Fig. 3 in Smith et al. (2016) (lines). Direct downwards fluxes are shown in red, diffuse are shown in black and the net flux (with upwards flux subtracted) is indicated in blue. In the legend, DFD is Direct Flux Downward (i.e. the solar beam), FD is Flux Downward (i.e. the diffuse downward propagating flux).*

Our results recreate nicely the net flux received by a flat surface. The direct flux downwards is, however, slightly overestimated and the diffuse flux slightly underestimated, although the total is in good agreement. Overall the model satisfactorily recreates the results of Smith et al. (2016).

### 2.1.3 Modelling Method

Zonal mean radiative transfer models were run for single sols at 5° intervals in *Ls* (0-355° inclusive) and latitude (-90-90° inclusive). The sun-path across the sky over the course of any given sol was calculated using an implementation of the Mars24 algorithm by Allison and McEwan (2000). Ninety-six timing points per sol were inputted to the Mars24 algorithm and only those points with zenith angles ($\theta_z$) less than 90° were taken as input to the D&A code.

The incident Solar flux at the top of the atmosphere was calculated for each band. The Solar 2000 ASTM Standard Extraterrestrial Spectrum Reference E-490-001 at zero air-mass was integrated between the wavelength limits of each band to give an in-band flux at zero air-mass at 1 AU. This was then scaled to the distance of Mars at the relevant Solar longitude (*Ls*).

Aerosol optical depths were taken from the Mars Climate Database scenarios for Mars Year 25-32 inclusive - complete-coverage reconstructed maps based upon TES observations (Montabone et al. 2015). MY 24 was not utilized as it was a partial-year dataset. Values were averaged across Martian longitude to give zonal mean optical depths as a function of latitude and *Ls*. These were verified against the published zonal mean maps. The optical depths obtained directly from the Mars Climate Database scenarios are for absorption only at 9.3 µm. These were multiplied by 2.6 to give the total extinction optical depths at ~880nm, as per Montabone et al. (2015). The zonal mean extinction optical depths as a function of latitude and *Ls* were averaged across all Mars Years to give a mean Mars Year map, shown in Fig. 2.

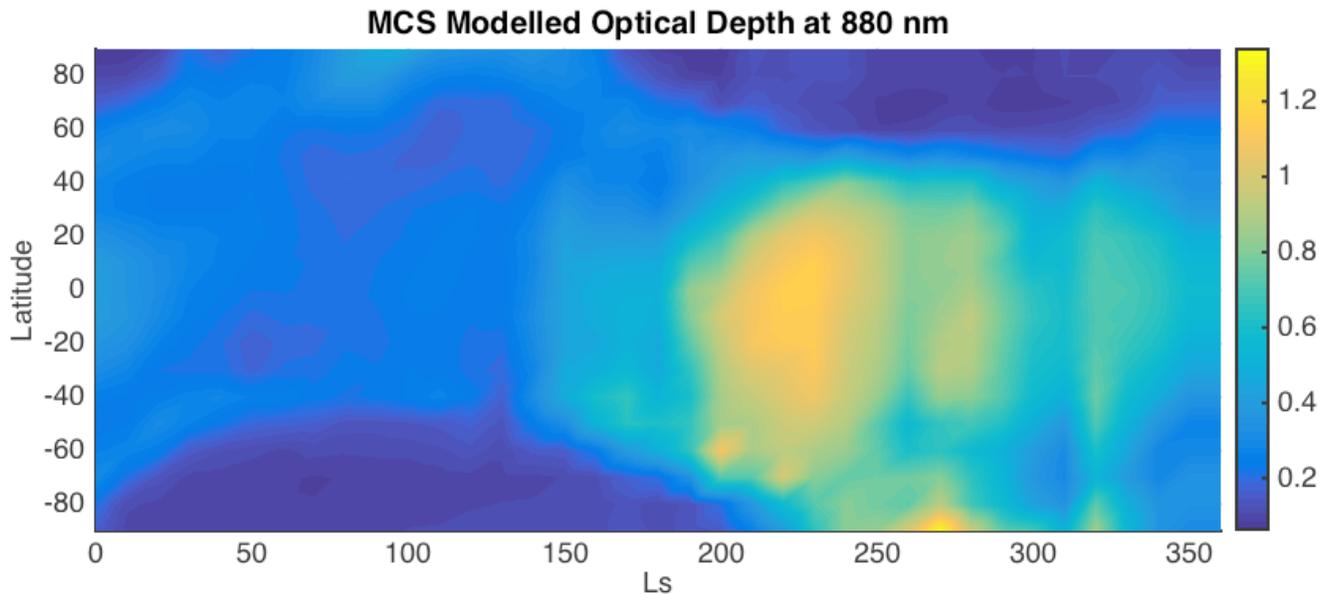

***Figure 2*** *– Map of zonal mean dust extinction optical depths as a function of Martian latitude and Solar longitude at 610 Pa.*

A model was constructed for 880 nm using the Mie parameters of Wolff et al. (2009). These were determined also assuming cylindrical particles with a 1:1 diameter to length ratio with an effective radius of 1.4 μm and effective variance of 0.3, but with wavelength dependent properties (refractive indices, single scattering albedo etc.) determined from visible and IR observations rather than the UV wavelengths of Wolff et al. (2010). Where comparisons were possible (e.g. single scattering albedo determined at 420 nm) there was good agreement between the two. The abundance of Mie scattering centers required to produce this optical depth at 880 nm at each point in latitude and *Ls* for the mean Mars Year were then computed. These Mie scattering center abundances were then inputted directly to the UV model across all bands, giving slight variation in dust optical depth as a function of wavelength.

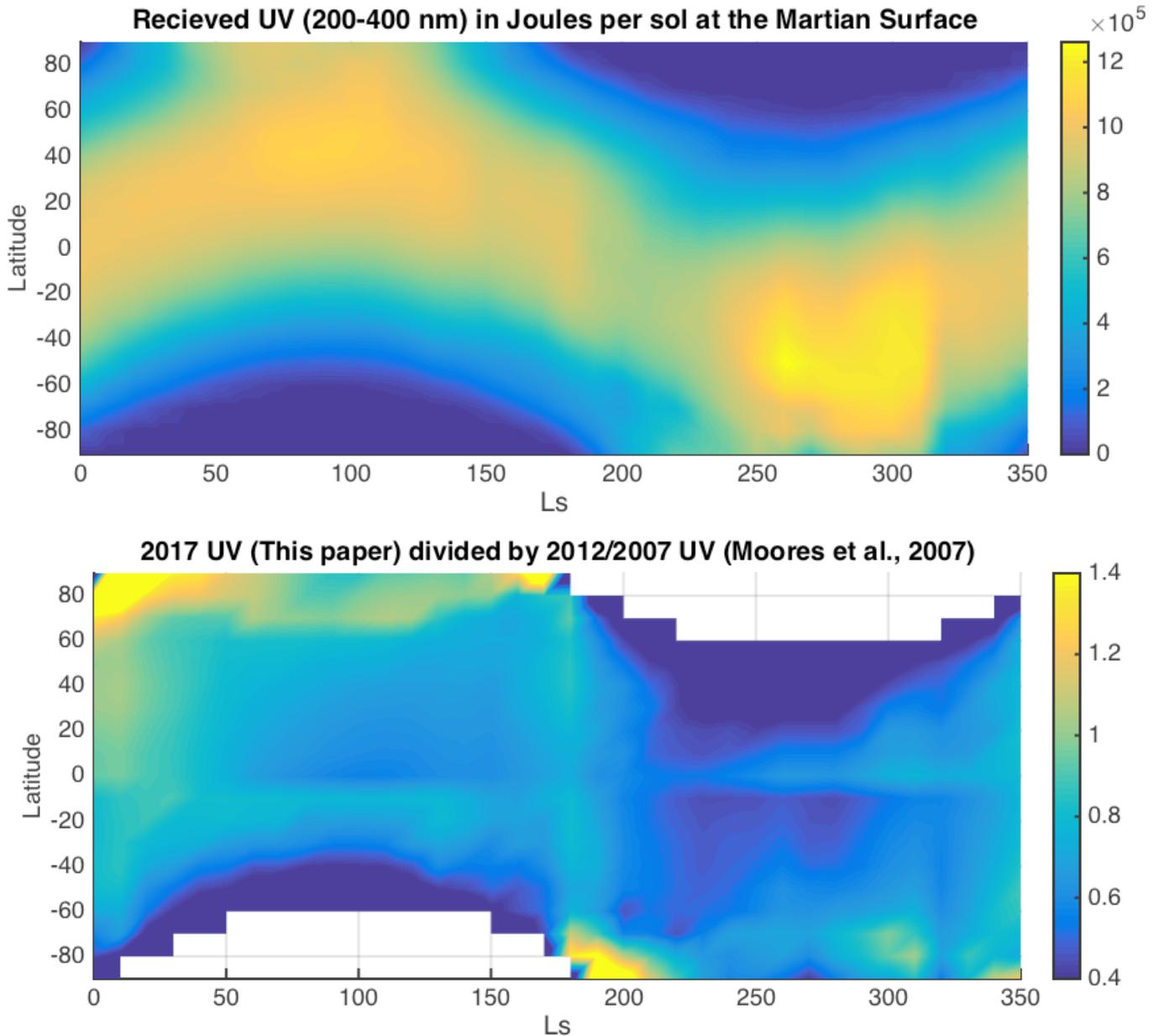

**Figure 3** – *(top panel) seasonal and latitudinal pattern of received UV energy per sol at the Martian surface as determined by the D&A code as validated using the REMS observations of UV from the MSL landing site at 4.5ºS (Smith et al., 2016). The annual variation in ozone and dust opacity from MCS retrievals (Figure 2) can be seen overprinting the approximately sinusoidal pattern of insolation resulting from the seasonal cycle in the distance of Mars from the sun and the orientation of the planet's spin axis, relative to the sun. (bottom panel) While there is some variability with season and latitude, a comparison of the REMS-validated 2017 UV Model to the 2012 UV Model shows that values from the model described in this paper are lower than pre-measurement estimates by 35%, on average.*

The D&A code was run over every point in the sun-path for each sol at each latitude and solar longitude, and the net downwards flux on a flat surface at each point calculated. The total energy received by a flat 1 m$^2$ area of the surface at a given latitude over a single sol at a particular *Ls* was calculated by integrating (trapezoidal approximation) the instantaneous net downwards fluxes from the D&A code when the solar zenith angle was less than 90º. For some sols, only one point in the sun-path was above the horizon and had a solar zenith angle close to 90º, so the total received energy on those sols was assumed to be zero.

The variation in sol-integrated UV energy produced by this model is shown in the top panel of Figure 3. These values are compared to the pre-REMS radiative transfer modeling work of Moores and Schuerger (2012) that derived values from Moores et al. (2007), which used a fixed optical depth of 0.5, in the lower panel of Figure 3. On average, the D&A code described here produces energies that are lower than the previous model by 35%. The two models agree best near the equator and diverge most strongly near the poles.

## 2.2 Refinement of the UV-CH$_4$ model

Using the updated model of UV flux at the surface of Mars, as described in section 2.1, the results of Schuerger et al. (2012) and Moores and Scheurger (2012) were refined. These refined results will be presented in section 3, while this section will review the UV-CH$_4$ model that was used to produce values of methane produced from irradiation of IDPs on the surface (Schuerger et al., 2012; Moores and Schuerger, 2012) when coupled to the UV radiative transfer model (Moores et al., 2007), which provided energy fluxes reaching the surface. Note that throughout this section and the remainder of the paper, methane production rates will be stated in terms of ppbv sol$^{-1}$. These units illustrate the column abundance and are derived by dividing the number of moles of methane produced per sol per m$^2$ of surface by the total number of moles per m$^2$ in the atmospheric column above.

### 2.2.1 Photon-Limited Production of Methane

In the case of Schuerger et al. (2012) an evaluation of the photon-limited case is provided in which the CH$_4$ produced originates with fresh material that completely covers the surface and the production is entirely dependant on the photon flux. This case is instructive as it sets an upper (if somewhat implausible) bound for methane production per sol. In order to determine the amount of methane emitted under these conditions, the quantum efficiency ($QE$) of the conversion of IDP organic carbon to methane is needed. Moores and Schuerger (2012) developed an equation for $QE$ as their equation 1 which captured the variation observed in methane flux under the experimental conditions described by Schuerger et al. (2012). This equation is reproduced below:

$$QE(\chi, T) = 2.76 \times 10^{-10} F_{CH4} \left[ 5.8 \times 10^{-4}(T - 273.15) + 0.126 \right] \left[ 6.9\,\chi + 0.026 \right] \qquad [\ 4\ ]$$

The $QE$ is provided here in mol J$^{-1}$, the methane flux in nmol g$^{-1}$ hr$^{-1}$ is $F_{CH4}$, the mass fraction of carbon is $\chi$ and $T$ gives the temperature in K. Though $F_{CH4}$ is a function of $\chi$ and $T$, each appears in this equation because it was observed by Schuerger et al. (2012) that as irradiation proceeded, an exponential decay was observed in the rate of methane production, likely due to a relative increase in more complex organics near the surface (e.g. kerogens) which meant that the surface organic carbon became progressively more resistant to photolysis. As such, this equation is written such that $F_{CH4}$ is a single fixed value at T= 298K and $\chi$ = 0.0169, appropriate for samples of the Murchison meteorite, and is selected from Figure 4 of Schuerger et al. (2012) to approximate the degree to which the decay in methane production had proceeded. The temperature and carbon content variables in equation 4 are then used to approximate how samples with different quantities of carbon at different temperatures would react to UV irradiation.

While Schuerger et al. (2012) considered cases in which this curve was integrated to simulate 120 sols of production, here only the limiting case is used in which samples are assumed to be freshly exposed and therefore $F_{CH4}$ = 0.145 nmol g$^{-1}$ hr$^{-1}$. Thus for Murchison at 298K and $\chi$ = 0.0169, a *QE* of 8.01 × 10$^{-13}$ mol J$^{-1}$ is derived and for IDPs with typical carbon contents of 10 wt% (e.g. Brownlee et al., 1985; Thomas et al., 1993) *QE* = 4.02 × 10$^{-12}$ mol J$^{-1}$ at 298 K. 100 wt% organic carbon is not a reasonable assumption, even for a limiting upper bound. While highly refined hydrocarbons can have organic carbon wt% in excess of 80%, amino acids more common to cometary sources, such as the glycine detected by the Stardust and Rosetta (Altwegg et al., 2016) missions, have carbon contents of up to ~30 wt%, comparable to the most carbon-rich IDPs retrieved from the stratosphere which have an organic carbon content of 24 wt% (Thomas et al., 1993). Thus, 30 wt% is taken as an upper limit with a *QE* of 1.18 × 10$^{-11}$ mol J$^{-1}$ at 298K. Under these two assumptions, keeping temperature as an independent variable, equation 4 is simplified to:

$$QE(\chi, T) = 4.87 \times 10^{-14} T - 2.73 \times 10^{-12} \qquad [\,5\,]$$

in the photon-limited case. Equation 5 will be combined with the modelled UV fluxes from section 2.1 and TES-derived average ground temperatures to determine the upper limit on the total production of methane per sol via the UV-CH$_4$ photolytic process. The total methane evolved in ppbv per sol, $N_{CH4}$, can be calculated according to:

$$N_{CH4} = 10^9 QE \times F_{UV} \frac{g_{Mars} M_{CO2}}{p_{CO2}}$$

[ 6 ]

Here, *QE* multiplied by the UV flux, $F_{UV}$, gives the total number of moles of evolved methane per m$^2$. This production rate is multiplied by the total number of moles of CO$_2$ into which this gas is mixed per m$^2$, in which $g_{Mars}$ is the surface gravity of Mars in N kg$^{-1}$, $M_{CO2}$ is the molecular weight of CO$_2$, taken to be 0.044 kg mol$^{-1}$, and $p_{CO2}$ is the surface pressure of CO$_2$, taken to be 610 N m$^{-2}$.

*2.2.2 Carbon-Limited Production of Methane*

In the case of Moores and Schuerger (2012), equilibrium values for organic carbon content of the surface are updated based upon known accretion rates of IDPs (Flynn, 1996) and the corresponding lifetimes of these particles under UV irradiation. As previously described by Moores and Schuerger (2012) lifetime calculations for individual carbon-bearing particles are more relevant for the conversion of this carbon to methane than are area-based calculations (e.g. Stoker and Bullock, 1998) in determining reasonable surface loads of organic molecules. The reason for this is that most photons that strike the surface do not interact with an organic carbon molecule. Hence this scenario is described as 'carbon limited' in contrast to the 'photon limited' situation of section 2.2.1 where all photons are assumed to interact with an organic carbon-bearing molecule. This carbon-limited approach represents a more realistic case that will allow global measurements of atmospheric methane, for instance those anticipated from TGO (e.g. Robert et al., 2016), to be linked to surface organic carbon content.

The analysis used in this paper mirrors the method described in Moores and Schuerger (2012) and the reader is directed to section 2.3 of this reference for the complete derivation of the particle lifetime, surface loading and methane produced. Adapted from Moores and Schuerger (2012) equations 3 to 7, the lifetime of a particle in size-bin $n$, $L_n$, where the size bins are those of Flynn (1996), the total surface mass of carbon, considering all size bins, $M_{surface}$, and the rate of methane production per sol from that surface loading of carbon, $N_{CH4}$, are given by:

$$L_n(\phi) = \frac{2 \chi D_n \rho f_{<900K}}{3 M_w F_{UV,AVG}(\phi) QE} \qquad [\,7\,]$$

$$M_{surface}(\phi) = \sum_n R_{n,accretion} L_n(\phi) \qquad [\,8\,]$$

$$N_{CH4}(L_S, \phi) = \sum_n \frac{R_{n,accretion} L_n}{\chi f_{<900K}} \frac{A_n}{m_n} F_{UV}(L_S, \phi) \cdot QE \qquad [\,9\,]$$

In these equations, $\chi$ is once again the carbon fraction, $D_n$ is the diameter in meters of particles in size bin n, $F_{UV,AVG}$ is the yearly-averaged UV flux in W m$^{-2}$, $M_W$ is the molecular weight in kg mol$^{-1}$, $\rho$ is the density of the particle in kg m$^{-3}$, the mass fraction of unaltered carbon for a single particle, $f_{<900K}$, the fraction of the accreting material heated to less than 900 K is derived from Flynn (1996). We take $QE = 2.76 \times 10^{-15} T - 1.54 \times 10^{-13}$, appropriate for a carbon content of 10 wt%, typical of IDPs, and the asymptotic flux of methane produced from extensively irradiated particles, $F_{CH4} = 0.024$ nmol g$^{-1}$ hr$^{-1}$. $R_{n,accretion}$ is the accretion of all particles in size bin n and has units of kg m$^{-2}$ s$^{-1}$ while $A_n$ and $m_n$ are the cross-sectional area and mass, respectively, of a single IDP particle in size bin $n$. Finally, $F_{UV}$ is the UV flux, a function of solar longitude, $L_S$, and latitude, $\phi$.

Equations 8 and 9, as written, assume that carbon-bearing particles, once accreted, do not move. However, such small particles are easily moved by the wind and have lifetimes that may approach 1000s of years (Moores and Schuerger, 2012). The alternative end member is that all particles are mixed via the atmosphere so that the surface concentration is the same everywhere. Thus the average value of $M_{surface}$ may be applied across the planet in equation 8. Results for both of these end members will be presented in section 3.3. Additionally, a redistribution is also considered where IDPs follow dust particles, accumulating in areas of enhanced dust abundance (Ruff and Christensen, 2002).

# 3. Results

## 3.1 Evolution of methane from concentrated surface sources

Using the photon-limited $QE$ and equation 6, the production of methane for a surface completely covered with organic carbon-bearing particles with 30 wt% organic carbon is shown in Figure 4. The peak of the plot is 3.9 ppbv sol$^{-1}$ which occurs during southern summer, just after perihelion in the region near 40ºS. If TGO or any other instrument makes observations of methane production at a rate above this value, the UV-CH$_4$ model of Schuerger et al. (2012) cannot be invoked to explain the result. In fact, even this upper limit case is extremely implausible as there exist few physical mechanisms of uncovering a large surface covered by completely fresh organic carbon on Mars. Furthermore, Schuerger et al. (2012) have shown that such high production rates cannot be maintained and will quickly fall after several sols as the material being irradiated degrades. Still, this case is instructive for understanding the maximum possible contribution that the UV-CH$_4$ process can make to the atmospheric column of methane on Mars.

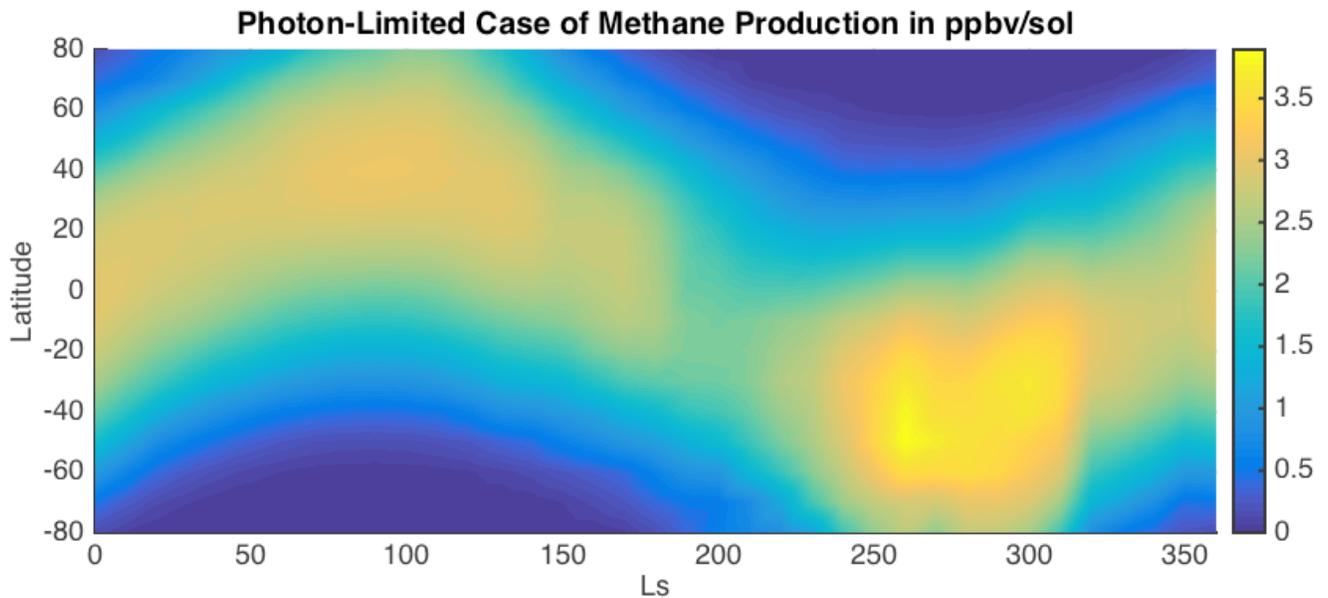

**Figure 4** – *upper limit ("photon-limited") case on the methane evolution rate in ppbv sol$^{-1}$ from the UV-CH$_4$ model assuming surfaces completely covered with freshly exposed 30 wt% organic carbon particles. Any methane evolution above these values that is observed (for instance, by TGO) is not consistent with the UV-CH$_4$ process and requires another production mechanism to be invoked.*

## 3.2 Concentration of Organics in regolith

Evaluating equations 7, 8 and 9 yields Figure 5, which describes the annual average UV flux on Mars, the lifetime of the IDP particles and the total amount of carbon mixed into the near surface. The predicted total UV flux by our model has fallen for Mars due to new REMS data (see section 2.1), and when compared to Moores and Schuerger (2012), yields longer lifetimes for individual IDP particles. With lower photolysis rates for the particles, more must accumulate before the total amount of input of carbon into the martian system from accretion is balanced with the total methane emission from the surface. Therefore the total amount of organic carbon in the soil must increase to produce the same amount of carbon loss to methane.

## 3.3 Evolution of methane from surface loading

The net result of this balancing is that the total emission of methane from the surface load remains nearly constant as compared to Moores and Schuerger (2012), as shown in Figure 6. Both end-member cases of complete redistribution of the IDP particles across the surface (6B) and no redistribution at all of these particles (6A) are plotted. As in the previous work (Moores and Schuerger, 2012), the peak emission of close to $2.9 \times 10^{-4}$ ppbv sol$^{-1}$ occurs with no redistribution and is concentrated in the polar regions, where there is the greatest difference in flux between winter and summer. With redistribution, it is the locations on the planet which see the highest flux where the greatest values of methane emission are found and here, the peak is somewhat lower at $2.0 \times 10^{-4}$ ppbv sol$^{-1}$.

Zonal variations are unimportant compared to meridional variations under either the complete redistribution or no redistribution models. However, if IDPs are mixed uniformly with surficial dust then they will tend to accumulate in areas with greater dust loading. As such, it is reasonable to consider weighting the distribution of IDPs according to the dust index derived and mapped by Ruff and Christensen (2002) using the MGS-TES instrument. The results of this analysis are shown as Figure 7 with the geographic production of methane examined for two times of the year: $L_S = 80º$ and $L_S = 260º$.

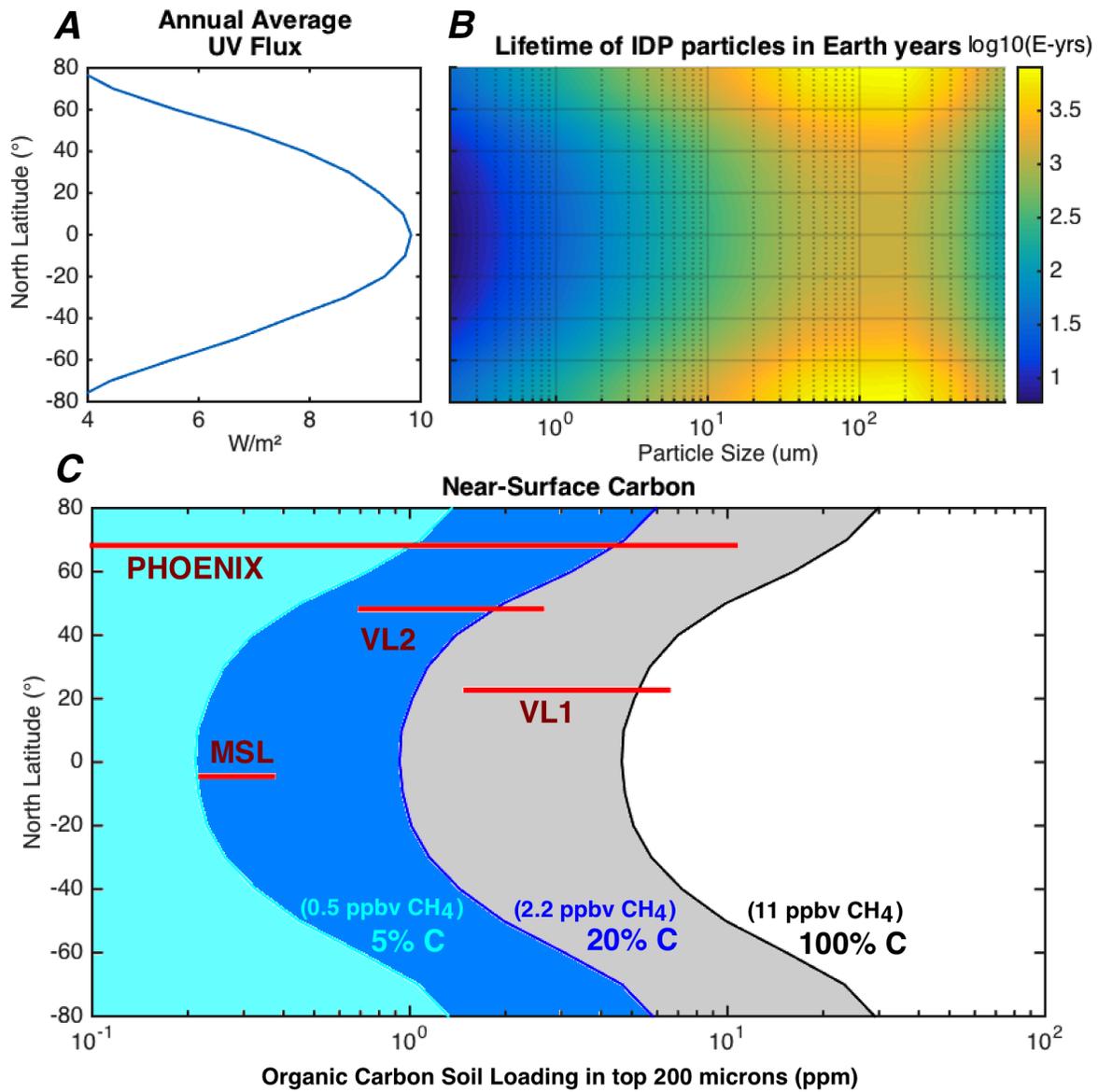

***Figure 5*** *– lifetime and surface concentration of organic carbon, assuming that destruction to methane is balanced by input into the system through accretion of organic carbon. The reduction in the UV flux (panel A) leads to longer lifetimes, measured in Earth Years (panel B) as compared to Moores and Schuerger (2012) and as a result larger surface concentrations of organic carbon in the surface regolith (Panel C). For panel C, the individual measurements of surface carbon derive from Archer (2010) for Phoenix, Navarro-Gonzales et al. (2011) for Viking and Freissinet et al. (2015) for MSL. The shaded areas of panel C indicates organic carbon concentrations in the soil consistent with mixing in the subsurface up to the equilibrium value for UV penetration into regolith of 200 µm. The different shaded regions describe different amounts of organic carbon (%C) in the system as compared to Flynn (1996) and how the surface and atmospheric concentrations are affected (see section 4.2 for a discussion of varying organic carbon input).*

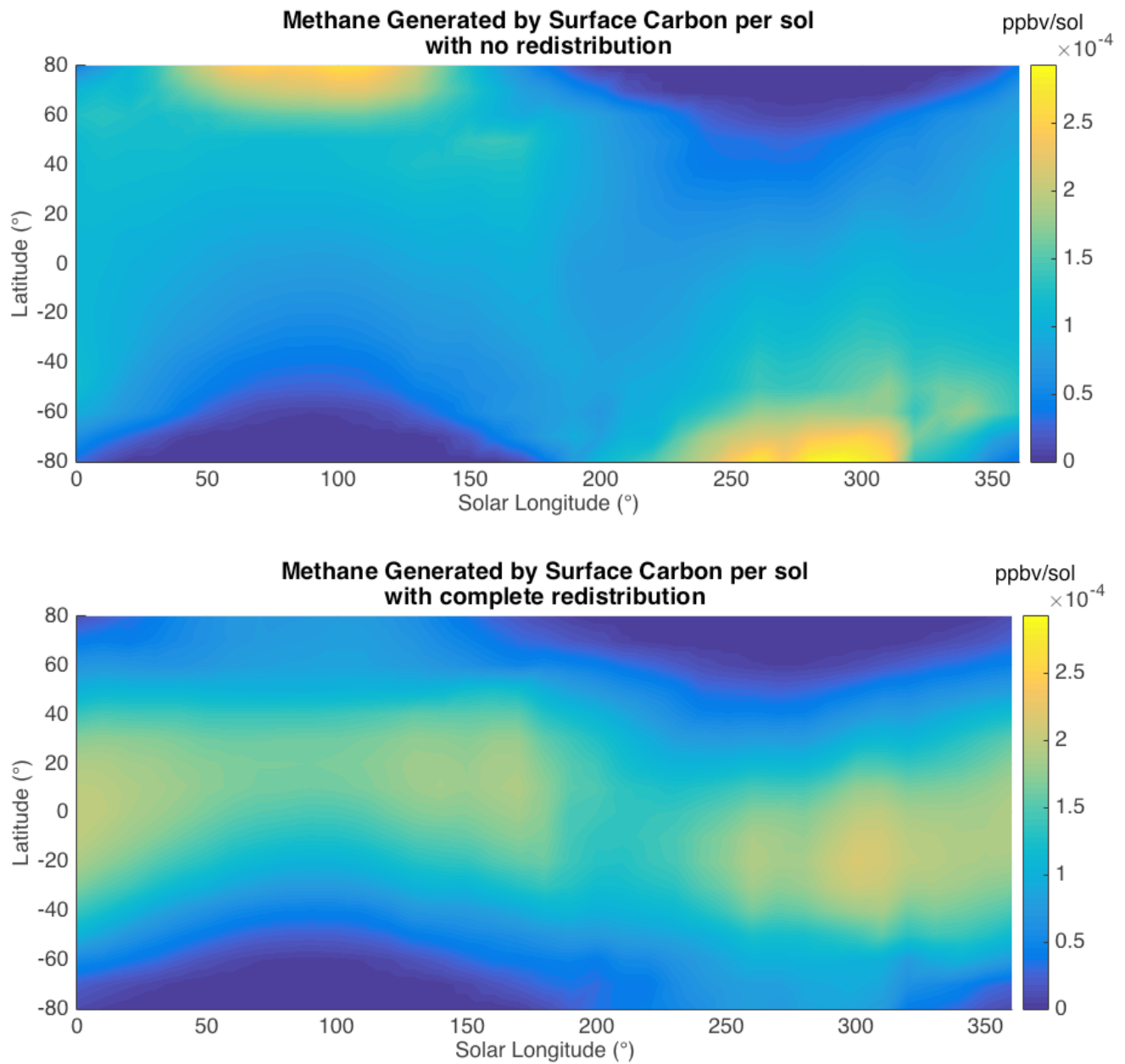

**Figure 6** – *methane produced by the UV-CH₄ process under two end members of IDP redistribution. At top, all IDPs are stationary once they arrive at the surface. Lifetimes at the poles are greater than at the equator, which leads to greater surface concentrations of organic carbon near the poles (e.g. Figure 5C). This accumulation in turn produces methane most rapidly near the solstices. At bottom, IDP concentrations are artificially homogenized across the planet, which leads to a peak in methane production where the peak daily UV energy is observed, closer to perihelion in the southern mid-latitudes.*

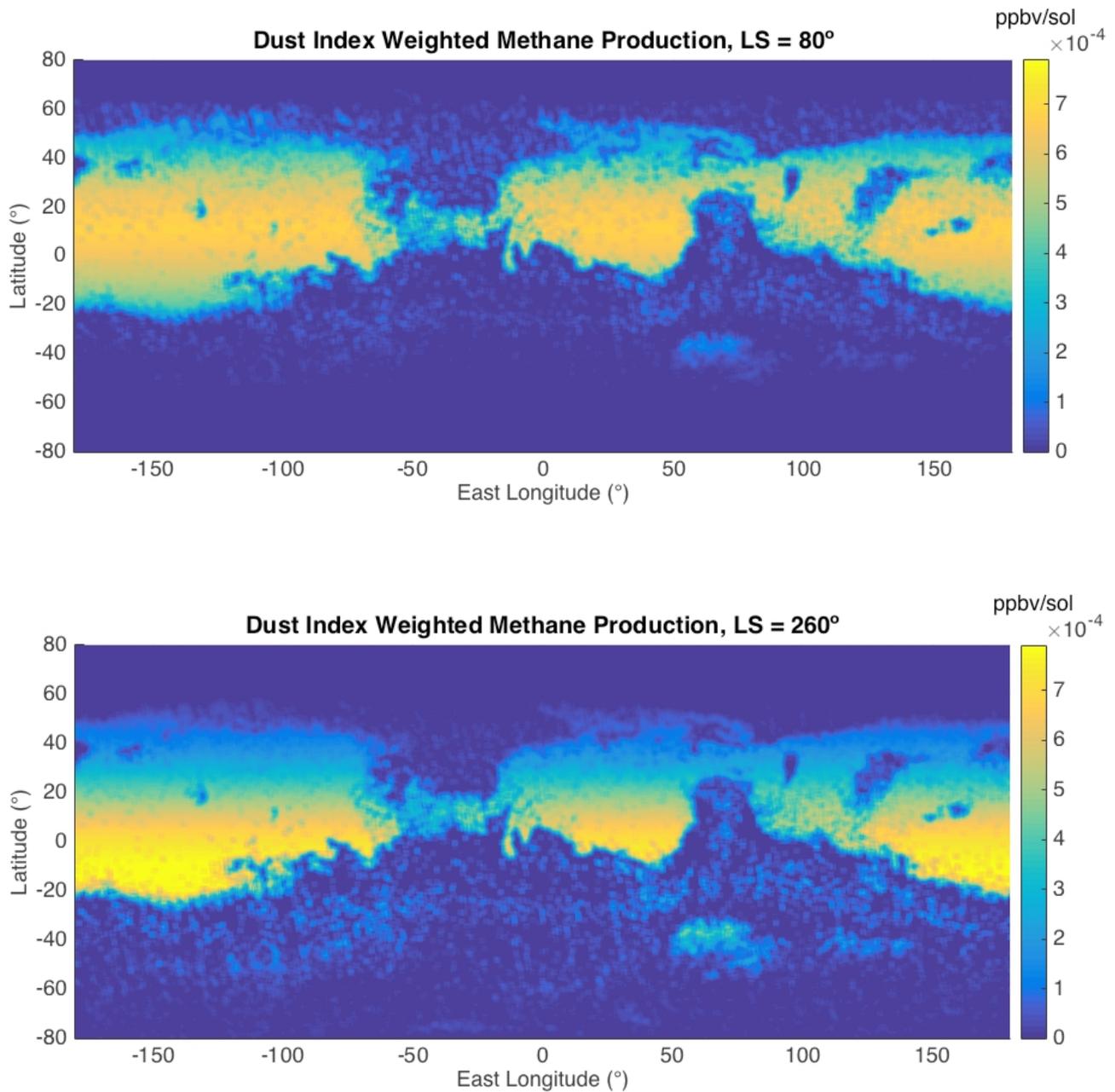

**Figure 7** – *geographic snapshots of methane production at two times of the year, $L_S = 80º$ (top) and $L_S = 260º$ (bottom). To produce these figures, IDPs were redistributed geographically based upon the dust index of Ruff and Christensen (2002) under the assumption that IDP particles and dust particles are equally mobile. Under this assumption, emission of methane in specific geographic areas can be larger than average values by a factor of up to ~2.7, with the greatest observed emission of $7.9 \times 10^{-4}$ ppbv $sol^{-1}$ in southern Tharsis.*

These two snapshots were selected to show the greatest range of production. At $L_S = 80º$, the maximum production is $6.8 \times 10^{-4}$ ppbv $sol^{-1}$, concentrated in the middle of the dusty band which lies between 20ºS and 50ºN and includes the two volcanic provinces, Tharsis and Elysium. At $L_S = 260º$, the peak in production is higher at approximately $7.9 \times 10^{-4}$ ppbv $sol^{-1}$. This peak is concentrated in the southern part of the dusty band and corresponds roughly to the peak in daily energy observed in Figures 3A and 6B. A much lesser peak is also observed in the Hellas basin.

In terms of the positions of spacecraft, Viking lander 1, Viking lander 2 and MSL are located on the edges of the dusty regions and the Phoenix Lander touched down in a relatively low-dust region. This could explain why the value for surface organics derived from Phoenix is not particularly elevated as compared to the other spacecraft, even though an elevated value would be expected due to the high latitude of the landing site. However, the variegated nature of the TES dust index near the other landed spacecraft make it challenging to conclusively attribute differences in the amount of organics detected to IDPs mixed into dust.

A final note of caution is required when interpreting the panels of Figure 7. First, the most common size fraction of IDPs is ~ 200 μm, much larger than the 1.0 μm modal diameter of dust particles. As such, IDPs may not be as mobile as are dust particles. Secondly, accumulating dust may bury accreted IDPs, as such, it is possible that the dusty regions shown in Figure 7 may contain more IDPs than do other areas, but the IDPs in dusty areas are protected from UV radiation and do not therefore emit methane at the same rate as IDPs in non-dusty areas. Finally, areas with the lowest dust index in Ruff and Christensen (2002) are not entirely dust-free regions, therefore the range of emission described in Figure 7 should be taken as a maximum range with emission likely closer to those shown in Figure 6B.

## 4. Discussion

### 4.1 Equilibrium Values of Methane in the SAM-TLS Era

Previous analysis of the accretion of organic carbon on Mars (Schuerger et al., 2012) from IDPs (Flynn, 1996) showed that there is sufficient accreting carbon to support 11 ppbv of methane in the atmosphere if all accreting carbon is converted into methane and the atmospheric lifetime of that methane is 329 years (Atreya et al., 2007). While the predicted value of 11 ppbv for the concentration of methane was broadly consistent with telescopic spectra of Mars that pre-dated the arrival of the MSL Rover, it seemed unlikely that the only product of the photolysis of IDP organic carbon would be methane. To constrain this branching ratio, which we define as the fraction of all IDP organic carbon that would become methane versus other products of destruction mechanisms, Schuerger et al. (2012) relied on the hydrogen to carbon ratio within IDPs of 20% to support a lower background value of 2.2 ppbv. Though it should be noted that more hydrogen is available within the martian environment at the surface (e.g. Meslin et al., 2013).

Moores and Schuerger (2012) extended this work, explicitly linking the UV flux and accretion of IDPs to the surface loading of organics and their photolysis to methane, describing a potential complete organic carbon cycle for Mars. With the UV flux now revised and validated by the first UV measurements at the bottom of the Martian atmosphere using the REMS instrument (Smith et al., 2016), the greatest change in the components of the UV-induced methane cycle is the total amount of organic carbon anticipated at the surface. This surface carbon reservoir made up of IDPs undergoing UV photolysis must increase as compared to the pre-MSL estimates of Moores and Schuerger (2012) as a direct result of the lower UV fluxes which extend particle lifetimes and slow the rate at which the UV-$CH_4$ process proceeds.

Conversely, the prediction of the total concentration of methane in the atmosphere is not changed by the current study. Under (1) the assumption of equilibrium in the Martian organic carbon cycle along with (2) estimates of total organic carbon input from Flynn (1996), (3) the branching ratio for conversion of organic carbon to methane versus other products and (4) the atmospheric lifetime of methane from Atreya et al. (2007), the concentration of methane is independent of the UV-$CH_4$ process and of the surface load of organics, so long as (5) the UV-$CH_4$ process is able to supply methane to the atmosphere on timescales shorter than the atmospheric lifetime. The range of reasonable values for each of these parameters, and the resulting effect on the change in observed methane concentration and surface organic carbon concentration, will be discussed in greater detail in section 4.2.

Yet, data from SAM-TLS shows that the methane concentration is never observed to be higher than 7 ppbv and can change quickly. Furthermore, the background value of methane is never greater than 0.7 ppbv (Webster et al., 2015). Both the magnitude of the methane concentration observed by SAM-TLS as well as its variability are surprising. Because the atmospheric lifetime is much longer than is either the vertical or geographic atmospheric mixing timescale, the observation of a relatively constant value of methane by SAM-TLS onboard MSL was anticipated. The downward revision of the UV flux described in this study cannot explain either of these behaviours. Still, the model described in the current study and in Moores and Schuerger (2012) provides a useful framework for considering how the SAM-TLS observations of methane might arise. In particular, either one of the assumptions making up the Moores and Schuerger (2012) description of the martian organic carbon cycle is incorrect or carbon is being removed by a competing, and as of yet unknown, process on Mars. Any such competing process must explain both the low overall value and the variability of methane in the martian atmosphere.

## 4.2 Where is the missing methane?

Examining this question requires consideration of each factor from section 4.1 which links the UV-$CH_4$ process to the total concentration of methane in the atmosphere. Factor (1), the assumption of equilibrium, is difficult to test. While it is known that Mars undergoes long-term changes in climate (Laskar et al., 2004) with sedimentation (Lewis and Aharonson, 2013) and exhumation (Kite et al., 2013) of dust deposits which should include admixed IDPs both taking place in different locations at different times in the past, it is unclear which process dominates in the current day. One large deposit that is known to be undergoing deposition is the North polar cap. Over the past hundred thousand years, the cap has added approximately 1 m of dust-rich ices (Smith et al., 2015). However, while this deposit may be able to trap some IDPs, it is unlikely to remove 75% of all accretion. As such, over the timescales appropriate to the methane atmospheric lifetime, equilibrium appears to be a reasonable assumption.

For factor (2), the carbon input to Mars through accreting IDPs, a simple solution to the disagreement between TLS measurements of < 0.7 ppbv and the model predictions for the background value of methane of ~ 2.2 ppbv is achieved if the total amount of carbon reaching the surface is smaller by a factor of > 3 as compared to the values listed in Flynn (1996), to < 80 tons of organic carbon per year. This could be the result of overestimates in the total IDP flux at Mars, in the organic carbon content of those IDPs or an underestimation in the fraction of carbon destroyed during atmospheric entry heating, perhaps due to discrepancies in the actual and predicted grain size distribution of IDPs accreting to Mars. Reducing the carbon input in this way not only reduces the atmospheric concentration of methane, but the amount of carbon anticipated to be mixed into the soil (Figure 5C).

Similarly, if the branching ratio, factor (3), is on the order < 7% instead of 20%, there is no disagreement between the model and observations. Recent work by Wadsworth and Cockell (2017), building on previous work by Shkrob et al. (2010), has argued that UV-activation of perchlorates in martian dust (Quinn et al, 2013) could result in enhanced destruction of microorganisms. While it is unclear whether such a process would yield methane, or if the process would degrade the organic carbon compounds found in IDPs, it is an example of a potential secondary process that could reduce the branching ratio to methane of available organic carbon at the surface of Mars. In addition to reducing the background methane concentration of the atmosphere, this factor also reduces the amount of organic carbon mixed into the soil by effectively reducing the amount of carbon available for conversion by the UV-$CH_4$ process. In the experiments of Schuerger et al. (2012), the total amount of methane produced was seen to correspond to a 5.5% conversion of organic carbon over the first 480 hours based upon estimates of the amount of carbon accessible to UV on these opaque surfaces derived from Jeong et al. (2003). This was extrapolated up to 20% over the entire lifetime of the particles, based upon the H/C ratios of the organic material (Schuerger et al., 2012). As such, the conversion to methane of the UV-CH$_4$ process is not likely to be less than 5% unless another process can consume the organic carbon more quickly than the UV-CH$_4$ process.

The different color-shaded regions of Figure 5C show the effect of changing the carbon input through either factor (2) or factor (3). As the total carbon available for the UV-CH$_4$ process is reduced, the surface concentration of organic carbon and the atmospheric concentration of methane decline together. Notably, the low atmospheric background of methane reported by Webster et al. (2015) is consistent with the low values of organic carbon observed in Gale Crater soils (Freissinet et al., 2014) indicating that the UV-CH$_4$ process remains a viable mechanism for producing the needed methane from this surface source.

Potential differences in factor (4) from the 329-year lifetime of Atreya et al. (2007) have been previously suggested by Lefèvre and Forget (2009) and are also implied by Fries et al. (2016). Shorter lifetimes also address lower background methane concentrations. However, unlike reducing the carbon input or the efficiency of the UV-CH$_4$ process, the surface concentration of carbon is not sensitive to atmospheric lifetime and will not change as the result of changes to this factor. A shorter methane lifetime of < 110 years would also bring model predictions into line with the SAM-TLS observations. With atmospheric lifetime, it is not possible to examine the total concentration separately from the variability of the methane concentration. In fact, as suggested by Lefevre and Forget (2009), lifetimes of atmospheric methane of decades or centuries are difficult to reconcile with the rapid disappearance of large amounts of methane over 120 sols, as telescopically observed by Mumma et al. (2009). They are also inconsistent with the spikes in methane to 7 ppbv observed by TLS (Webster et al., 2015) if the sources producing these spikes are large.

However, if the sources producing methane at Gale crater are local and small, their rapid disappearance may be explained by mixing with the atmosphere at large, as demonstrated by the lagrangian tracer mesoscale atmospheric models of Moores et al. (2016). The initial rise, however, is unlikely to result from photolysis of surface organics described within the UV-$CH_4$ model, due to factor (5). To provide an example, 14 sols separate the 0.56 ppbv measurement on sol 292 ($L_S$ = 328.6) and the 5.78 ppbv measurement on sol 306 ($L_S$ = 336.5) (Webster et al., 2015). Examining Figure 4, maximum UV-$CH_4$ production rates at this time of year and latitude are ~3 ppbv sol$^{-1}$, which means that it would take two sols to produce the needed methane. This amount of time is larger than the mixing timescale of the crater of less than a sol (Moores et al., 2016) which makes it difficult to contain that much methane despite Gale being a relatively isolated system (Moores et al., 2015; Tyler and Barnes, 2013; Rafkin et al., 2016; Pla-Garcia et al., 2016). Furthermore, the production of methane from surface organics would require exceptionally rich materials (> 20 wt% organic C) to have been suddenly and quickly uncovered over a large surface area. As such, while this scenario is possible, at the limit of the UV-$CH_4$ model, it is not plausible. Thus, while the refined UV-$CH_4$ model in this work can provide sufficient methane quickly enough (factor 5) to explain the background values, it cannot explain the spikes in methane concentration.

**4.3 Production and Destruction of Methane during atmospheric passage**

An alternative explanation for the spikes has been suggested by Fries et al. (2016), who suggest that methane spikes could be explained by an enhanced UV-$CH_4$ process working near the top of the martian atmosphere as seasonal enhancements in the IDP flux, termed meteor streams, are subjected to photolysis. Here, fluxes of UV radiation are somewhat higher than they are at the surface, which would allow methane to be created more quickly. Simultaneously, greatly enhanced lyman-$\alpha$ flux would allow the produced methane to also be destroyed much more quickly (Wong et al., 2003). Such an explanation is attractive as it provides a means for explaining the highly variable nature of the reported methane results, none of which agree in geographic extent or timing (e.g. Krasnopolsky et al., 2004; Formisano et al., 2004; Geminale et al., 2008; 2011; Mumma et al., 2009) without concluding that all results are therefore suspect (Zahnle et al., 2011).

Fries et al. (2016) contend that the timing of meteor streams correspond to all known spikes of methane. However, Roos-Serote et al. (2016) have challenged the timing of the Fries et al. (2016) mechanism, while the mass contained within the meteor streams described by Fries et al. (2016) appears to be too small to form the methane spikes. In fact, no methane spike was observed by TLS even from the historically close passage of Comet Siding Spring in 2014, in which Mars passed directly through a cometary coma that deposited more than an order of magnitude more IDP-like dust, between 2700 and 16,000 kg (Schneider et al., 2015), than the largest of the Fries et al. (2016) meteor streams into the upper atmosphere of Mars over the course of a few hours. This value is supported by total electron column densities in the meteoritic layer of the martian atmosphere of 2 x $10^{15}$ e$^-$ m$^{-2}$ (Restano et al., 2015; Gurnett et al., 2015) as compared to typical peak column densities of 3-4 x $10^{14}$ e$^-$ m$^{-2}$ (Pandya and Haider, 2012).

While it is possible that Siding Spring-derived materials have a larger ablated fraction as compared to typical meteoritic accretion due to increased velocity (Whithers, 2014), it seems unlikely that any of the hypothetical streams described by Fries et al. (2016) could deposit more than 1 x $10^4$ kg of dust into the martian atmosphere, which does not compare favourably to the 1.9 x $10^7$ kg of methane observed by Mumma et al. (2009). A robust upper limit on methane produced from Siding Spring dust may be obtained by using the model of Moores et al. (2014). Estimates of cometary dust production at encounter (October 19, 2014) from Opitom et al. (2016) indicate an *Afρ* of 1020 cm at a wavelength of 445.3 nm and cometocentric distance of 10 000 km. This data can be combined with the relationships of A'Hearn et al. (1995) and Moorhead et al. (2014) to derive a total absolute magnitude of 9.79 at encounter. At this magnitude, the model of Moores et al. (2014) would predict a spike in martian atmospheric methane of no more than 1 $\times 10^{-5}$ ppbv, even for the most optimistic assumptions. This represents a level of methane enhancement unobservable to any current measurement technique.

Despite these challenges, the idea put forward by Fries et al. (2016) of photolysis during atmospheric passage is worthy of consideration. If a large fraction of methane is deposited high in the atmosphere, the consumed organic carbon becomes unavailable to surface reservoirs, thereby reducing the equilibrium atmospheric value. Furthermore, small particles may be lofted for long periods of time, in some cases for timescales comparable to their UV-CH$_4$ lifetimes, based on Stokes settling rates for suspended particles (Melosh, 2011). Therefore, it would be helpful to clarify where in the atmosphere the particles undergoing photolysis deposit methane, as both Fries et al. (2016) and Roos-Serote et al. (2016) conflate measurements of methane with TLS at the surface and telescopic measurements of the entire atmospheric column which sample different ranges of atmospheric altitude. Furthermore, TGO will sample the full column of the atmosphere in nadir mode along with retrieving vertical profiles in occultation (Robert et al., 2016). The sources and implications of the methane results so-obtained are different depending on whether the methane is found predominantly high or low in the atmosphere.

In order to create a model of photolysis while suspended, the particles were treated as spherical bodies and their velocities through the atmosphere treated using Stokes settling protocols, as modified for the Knudsen number following the methods of Taylor et al. (2007) and Murphy et al. (1990). For details of this method, including the relevant equations, see Moores et al. (2016), section 2.2.1. The pressure and density profile of the atmosphere was approximated as an exponential decline from a surface pressure of 610 Pa with a typical pressure scale height of 10.7 km, appropriate for a hydrostatic isothermal $CO_2$ atmosphere at 210 K. In the model, the particles enter the top of the martian atmosphere at hypersonic velocity and decelerate very rapidly, achieving terminal velocity at high altitude. This altitude, typically between 80-105 km, is approximated by using the layer of metal ions observed in the Martian atmosphere that are thought to be derived from larger IDPs ablating surface materials (e.g. Pandya and Haider, 2012; Gurnett et al., 2015). As such, 100 km is used as the starting altitude for methane generation for the IDPs. Fluxes of UV to create methane were assumed proportional to the optical path above the particles, approximating a constant mixing ratio with height, with the overall optical depth of the atmosphere fixed to an average value of 0.5.

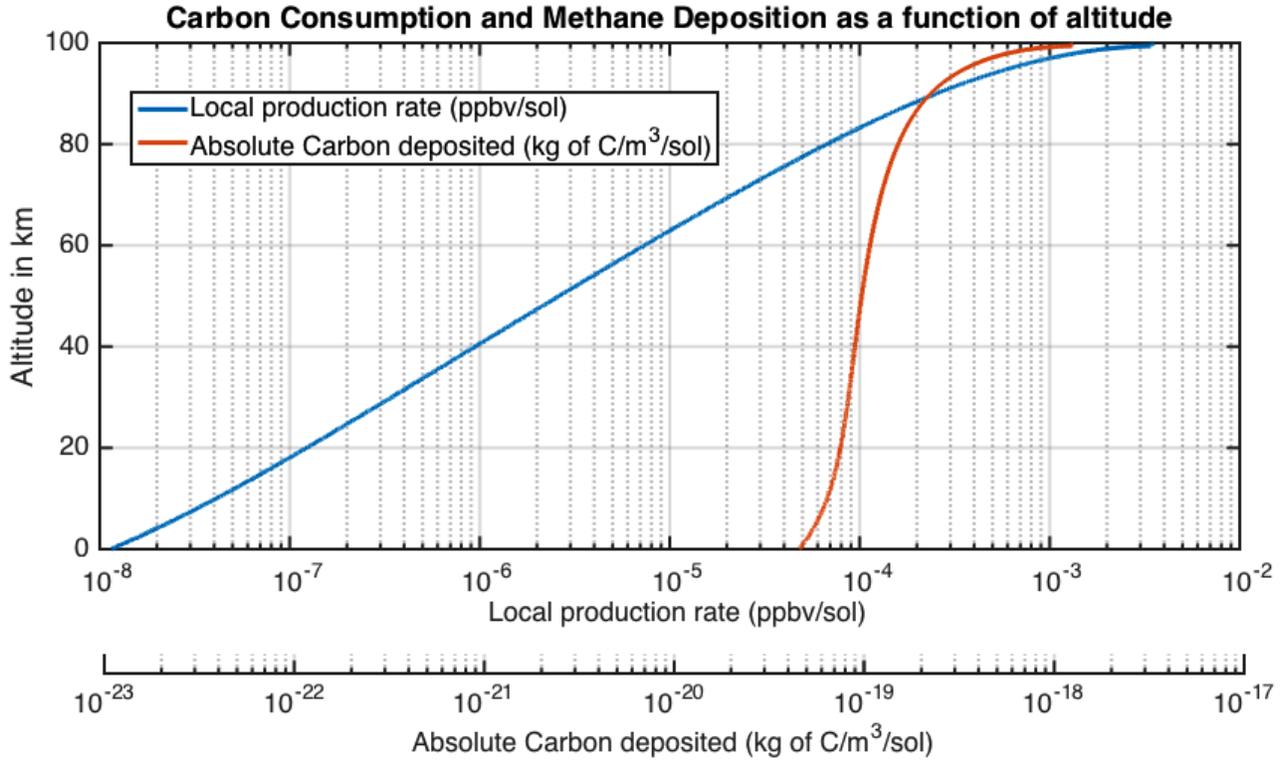

**Figure 8** – *Absolute Carbon Deposition in kg $m^{-3}$ $sol^{-1}$ (Red) and its conversion to local methane production in ppbv $sol^{-1}$ (Blue) resulting from sedimenting IDPs as a function of altitude. Relatively more carbon is deposited at high altitude resulting in relatively high local methane production rates, when the small number density of admixed $CO_2$ at these altitudes is taken into account. Overall, the total column-integrated carbon deposition is 1.4 x $10^{-14}$ kg $m^{-2}$ $sol^{-1}$ and the total column methane production produced by dividing this by the total mol of atmosphere in the column is 8.7 x $10^{-8}$ ppbv $sol^{-1}$. This latter number can be directly compared to surface production as both express methane production compared to the total atmosphere. Atmospheric sources from sedimenting particles are therefore much less important than surface sources.*

The results of the sedimentation of particles are shown in Figure 8. The total amount of carbon consumed by the UV-$CH_4$ process and deposited in the atmosphere as the particles fall is shown in red in units of kg of carbon per cubic meter per sol. In all, the carbon deposited into the atmosphere integrates to $1.4 \times 10^{-14}$ kg m$^{-2}$ sol$^{-1}$. Once the volumetric concentration of carbon is converted into moles of methane per cubic meter and divided by the molar density of the atmosphere, a concentration production curve, given in ppbv per sol, is derived and shown in blue. While the local values at high altitude appear to be an order of magnitude above surface production rates (Figures 6 and 7), the integrated column production of methane is only $8.7 \times 10^{-8}$ ppbv sol$^{-1}$. In the interpretation of this value, note that the majority of atmospheric $CO_2$ is located within one scale height (10.7 km) of the surface, as such, while concentrations are relatively high at high altitude, they represent a relatively small amount of methane and admixed atmosphere. Furthermore, recall that surface-produced methane was assumed evenly mixed with the entire atmosphere in order to produce concentration and production values. The increasingly rapid tail-off of both curves (though more apparent in the Red carbon deposition curve) in the lower few scale heights of the atmosphere is the result of increasing atmospheric opacity.

How does the atmospheric production compare to surface production? For the smallest particles described by Flynn (1996), photolysis lifetimes are short compared to the amount of time these particles may be held aloft. For diameters of 2 μm and smaller, the particles are smaller than the average atmospheric dust and atmospheric settling times are effectively infinite. These small particles are therefore entirely consumed in the atmosphere with larger particles consumed relatively closer to the surface. However, these small particles contain only 0.14% of all IDP-derived carbon delivered to Mars (Flynn, 1996) making their contribution to the overall carbon budget relatively inconsequential. As the particle size increases, the fraction of the particle lifetime spent in the atmosphere decreases rapidly. Over all sizes, only 0.32% of all IDP-derived carbon is consumed in the atmosphere, with the remainder making it all the way to the surface, where it accumulates in the top 200 μm of surface materials until destruction and production are equal. It is for this reason that the surface carbon can play an outsized role in the diurnal methane production rate.

Even the high values of production at high altitude may not result in particularly high local concentrations. First of all, based on figures from Wong et al. (2003), rates of destruction of methane above 70 km are higher by orders of magnitude than at the surface. As such, the effective lifetime of methane above 70 km is likely to be low, perhaps less than a single Martian year. This means that high altitude methane cannot accumulate as easily as methane near the surface. Secondly, a key question, which remains unanswered, is whether interactions with the martian environment are needed to liberate material from IDPs that becomes susceptible to the UV-$CH_4$ process. Before arriving at the top of the martian atmosphere, IDPs have spent long periods of time subjected to higher fluxes of higher energy radiation (including UV radiation) in interplanetary space as they would receive on the martian surface. Yet it is known that these particles preserve significant quantities of organic carbon (Brownlee, 1985; Thomas et al., 1993). Moores and Schuerger (2012) hypothesized that it was the martian environment's ability, through physical collisions with other particles, to break apart IDPs, which are composed of clumps of friable nm-sized particles held together with an organic glue (Flynn et al., 2010), which made their interiors susceptible to UV photolysis.

## 4.4 Implications for the Trace Gas Orbiter (TGO)

The analysis of section 4.3 indicates that most IDPs make it to the surface with their organic carbon intact. As such, if IDPs are the source of the atmospheric methane observed on Mars, it may be assumed that methane is being injected into the atmosphere at the surface. That surface reservoir may vary geographically and therefore, there will be geographic variations in methane production rates. If IDPs cannot move, production will be concentrated near the poles (Figure 6A). If the IDPs are uniformly mixed across the surface by aeolian processes, then peak production will occur near the southern mid latitudes (Figure 6B), where daily-integrated energy values are highest (Figure 3). Finally, if IDPs are associated with reservoirs of dust, the greatest production will be seen in southern Tharsis (Figure 7B) near Perihelion. By deconvolving these surface sources, TGO may therefore be able to constrain the concentration of surface organic carbon across the planet.

However, such a deconvolution will be challenging and would therefore rely on modelling of the movement of methane in the martian atmosphere (e.g. Mischna et al., 2011). Partly, this arises from the necessity of using the occultation mode to maximize the sensitivity of the NOMAD instrument, which may be as small as 25 pptv (0.025 ppbv) for a single measurement or 10 pptv (0.01 ppbv) for averaged spectra (Robert et al., 2016). Note that the nadir mode, which has a relatively small geographic footprint, is limited to a sensitivity of 11 ppbv (Robert et al., 2016) and is not useful for correlating surface carbon to atmospheric methane via the UV-CH$_4$ process. More significantly, the low production rates described in the current study of less than 0.001 ppbv sol$^{-1}$ mean that the atmosphere will be able to mix methane over a broad geographic area on timescales of tens to hundreds of sols needed for the production to build up to observable levels. The 300 m expected vertical resolution of the NOMAD instrument (Robert et al., 2016) should help discern the source of variations in methane production as geographic and vertical mixing proceeds at different rates on Mars.

Finally, in the event that TGO observes large plumes of methane, if these are seen to increase at more than a few ppbv per sol, the UV-CH$_4$ process is ruled out for plume formation (Figure 3). Such rapid production of methane necessitates a substantial source of methane, such as the subsurface release discussed by Mumma et al. (2009).

## 5. Conclusions

The model of Moores and Schuerger (2012) is updated here to account for the first ever measurement of UV fluxes from the surface of Mars by the REMS instrument. In terms of daily integrated energy, the revised values presented in this paper are lower than the 2012 modeled values by ~35%. The reduction in UV to drive the UV-CH$_4$ process (Schuerger et al., 2012) leads to an increase in the lifetime of IDPs accreted by Mars as well as the mass of organic carbon needed at the surface to balance methane emission with organic carbon accretion. However, the rate at which this surface loading of IDP organic carbon releases methane into the martian atmosphere is unchanged from 2012 values (Moores and Schuerger, 2012) and the amount of surface organic carbon remains consistent with derived values from all landed spacecraft. In particular, the low concentration of background atmospheric methane (Webster et al., 2015) and the low concentration of organics in the soil (Freissinet et al., 2014) may be consistently linked through the UV-$CH_4$ process.

In addition to the two cases discussed in previous work of no redistribution and total redistribution of IDPs after accretion by Mars (Moores and Schuerger, 2012), the current study adds a third case in which IDPs accumulate where dust is found. This results in an increase of the maximum emission rate of methane of over a factor of two, from $2.9 \times 10^{-4}$ ppbv sol$^{-1}$ to $7.9 \times 10^{-4}$ ppbv sol$^{-1}$, localized to the southern part of Tharsis, near perihelion ($L_S = 251^\circ$).

Background values for methane of < 0.7 ppbv as measured by Curiosity (Webster et al., 2015) are much lower than were predicted by Schuerger et al. (2012) of 2.2 ppbv. However, these equilibrium values can still be described within the UV-$CH_4$ framework. Possible causes of this discrepancy were considered. Either (1) the organic carbon cycle on Mars is not in equilibrium with most organic carbon accreted being actively buried, (2) accretion of organic carbon to Mars is smaller than predicted (Flynn, 1996), (3) other reactions aside from the UV-$CH_4$ process reduce the amount of organic carbon available for photolysis or (4) the atmospheric lifetime of methane is smaller than previously calculated (Atreya et al., 2007). Of these four factors, the discrepancy between SAM-TLS background values and past work (Moores and Schuerger, 2012; Schuerger et al., 2012) would be eliminated if the accretion of organic carbon reaching the surface were reduced from 240 tons to < 80 tons per year, if the atmospheric lifetime of methane is < 110 years, if the efficacy of the UV-$CH_4$ process is reduced from 20% to < 7% or by any combination of these factors which reduce the amount of carbon cycling through methane in the martian environment by a factor of > 3. Disequilibrium was judged to be less likely than the other three factors. Furthermore, the spikes of methane observed by SAM-TLS are not explainable via the UV-$CH_4$ process, as too much methane must be produced in too short a time.

Finally, the mechanism of Fries et al. (2016) whereby IDPs are photolyzed high in the atmosphere where methane destruction rates are faster than at the surface was tested. Of the carbon deposited into the atmosphere during sedimentation, the majority is deposited at high altitude. However, the total atmospheric deposition of methane by sedimenting particles was limited to 0.32% of all organic carbon delivered to Mars, resulting in column methane production rates of no more than 8.7 x $10^{-8}$ ppbv $sol^{-1}$. As such, atmospheric deposition during sedimentation does not represent a significant source of methane and cannot affect in a significant way the equilibrium value of methane nor the large excursions in methane concentration observed by SAM-TLS or telescopic spectroscopy.

## 6. Acknowledgements

JEM was supported in this research by a Natural Sciences and Engineering Research Council of Canada (NSERC) Discovery Grant (436252-2013). CLS was supported by a fellowship under the Technologies for Exo/Planetary Science (TEPS) Collaborative Research and Training Experience (CREATE) programme of NSERC. ACS was supported by a NASA grant (NNX14AG45G) from the Mars Fundamental Research program. The authors would like to particularly thank Lyn Doose of the University of Arizona for assistance with compiling the D&A algorithm.